\documentstyle[multicol,aps]{revtex}

\begin{document}
\newcommand{\la} {\langle}
\newcommand{\ra} {\rangle}


\title{Enstrophy and dissipation must have the same scaling exponent in the high Reynolds number limit of fluid turbulence}

\author{Mark Nelkin$^{1}$ }

\maketitle

\begin{abstract}

Writing the Poisson equation for the pressure in the vorticity-strain form, we show that the pressure has a finite inertial range spectrum in the high Reynolds number limit of isotropic turbulence only if the anomalous scaling exponents $\mu$ and $\mu_{\omega}$ for the dissipation and enstrophy (squared vorticity) are equal.  Since a finite inertial range pressure spectrum requires only very weak assumptions about high Reynolds number turbulence, we conclude that the inference from experiment and direct numerical simulation that these exponents are different must be a finite range scaling result which will not survive taking the high Reynolds number limit.

\end{abstract}

\bigskip

\noindent $^{1}$Physics Department, New York University, New York, NY 10003,\\
and Levich Institute, CCNY, New York, NY 10031 USA\\
Electronic address: Mark.Nelkin@nyu.edu
\begin{multicols}{2}
\narrowtext

It has long been recognized \cite{obuk} that the local rate of dissipation in high Reynolds number turbulence exhibits anomalous scaling, characterized by the fluctuations in $\epsilon_r({\bf x},t)$, the locally averaged dissipation rate over a region of size $r$, where $\epsilon$ is the local rate of dissipation per unit mass of turbulent kinetic energy.  Here we find it more convenient to work with the spectrum of dissipation fluctuations in the inertial range, which is conventionally assumed to have the form

\begin{equation}
E_{\epsilon}(k) = C_{\epsilon} \la \epsilon \ra^2 k^{-1} (kL)^{-\mu}, 
\label{diss}
\end{equation}
where $\mu$ is the putatively universal exponent describing the dissipation fluctuations in high Reynolds number turbulence \cite{monin},$ \la \epsilon \ra$ is the average rate of energy dissipation per unit mass, $k$ is the wave number, $L$ is the geometrically defined external length scale, and $C_{\epsilon}$ is a dimensionless constant which is probably not universal.

For later purposes, note that the dissipation rate $\epsilon$ is  
related to the symmetric part of the strain tensor as
\[ \epsilon = 2\nu S^2, \]
where $S^2 = S_{ij}S_{ji}$, $S_{ij}=1/2(\partial u_i/\partial x_j  
+ \partial u_j/\partial x_i)$, and $\nu$ is the kinematic viscosity of the fluid. The relationship between $S$,   
the antisymmetric part of the strain  
tensor $\sigma_{ij}= 1/2(\partial u_i/\partial x_j - \partial  
u_j/\partial x_i)$ and the enstrophy $\Omega \equiv {\bf  
\omega}^2 $, is 
\begin{equation}
S^2 - \frac{\Omega}{2} = \frac{\partial u_i}{\partial x_j}
\frac{\partial u_j}{\partial x_i}.
\label{rela}
\end{equation}  
Using the Navier-Stokes  
equations,  (\ref{rela}) can also be related to  
the pressure $p$ via
\begin{equation}
\nabla^2 p = \frac{\Omega}{2} - S^2,
\label{pre}
\end{equation}
or

\begin{equation}
\nabla^2 p = - \frac{\partial u_i}{\partial x_j}
\frac{\partial u_j}{\partial x_i},
\label{press}
\end{equation}
and the density of the fluid is taken as $\rho=1$ with no loss of generality. The derivation of (\ref{rela}), (\ref{pre})  and (\ref{press}) makes essential use of incompressibility.

For the enstrophy, a quantity analagous to the dissipation can be defined as

\begin{equation}
D = \nu {\omega}^2.
\label{epsom}
\end{equation}
The anomalous scaling of D is exhibited in its power spectrum, which is assumed to have the inertial range form
\begin{equation}
E_D(k) = C_D \la \epsilon \ra^2 k^{-1} (kL)^{-\mu_{\omega}},
\label{spec}
\end{equation}
where $\mu_{\omega}$ is a scaling exponent describing the anomalous scaling of enstrophy fluctuations.

There has been an interesting recent controversy concerning the relative scaling of enstrophy and dissipation.  Older direct numerical simulations (DNS) \cite{sigg}, \cite{kerr} showed that vorticity was more intermittent than strain for single point quantities.  The one-dimensional measurements of the streamwise components of  
$\epsilon$ and $\omega$, obtained at both high and low Reynolds  
numbers \cite{mene}, conclude that the degree of intermittency  
in the dissipation and enstrophy fields are not the same.  On the other hand L'vov and Procaccia \cite{lvov} have argued on general symmetry grounds that the asymptotic scaling exponents must be equal; i.e. that $\mu_{\omega}=\mu$.  This motivated Chen, Sreenivasan and Nelkin \cite{vor} to study the inertial range scaling of enstrophy and dissipation in a DNS at moderate Reynolds number $R_{\lambda}=216$.  They found that the enstrophy was clearly more intermittent than the dissipation suggesting that $\mu_{\omega}>\mu$.  Finally, He et al \cite{he98} have studied  the statistics of dissipation and enstrophy induced by a set of Burgers vortices.  They found that finite-range scaling exponents are different for these two quantities, but that for this model system, the asymptotic scaling exponents for dissipation and enstrophy can be shown to be equal in the limit of infinite Reynolds number.  It is then of considerable interest to see if this can be shown more generally  without resort to a particular model system.

In this paper I show on quite general grounds, starting from (\ref{pre}), that $\mu_{\omega}=\mu$.  I use only the assumption that the pressure has a finite power spectrum in the inertial range when $\nu\rightarrow 0$.  The qualitative idea is that each of the source terms on the right hand side of (\ref{pre}) scales as $(1/\nu)$ but their difference must be independent of $\nu$.  This can only be true if each source term separately has the same scaling exponent.  

In writing (\ref{diss}) and (\ref{spec}), the idea of an inertial range has been mentioned, but it is important to define it more precisely.  The 
basic assumption is that certain quantites are independent of the kinematic viscosity $\nu$ in the limit that $\nu\rightarrow 0$.  This is conventionally assumed for the average dissipation rate $\la \epsilon \ra $, and the usual picture of dissipation fluctuations assumes that it is also true for the spectrum of dissipation flucutations.  This is indicated in (\ref{diss}) and (\ref{spec}) by the appearance of the extenal length scale $L$ rather than the dissipation length scale $\eta$.  It is conventionally assumed that the pressure spectrum has a finite inertial range.  This follows naturally from the Poisson equation in the form (\ref{press}).  Hill and Wilczak \cite{hill95} have shown that the pressure structure function $\la [p(x+r)-p(x)]^2 \ra $ can be written as an integral over the three independent fourth order velocity structure functions $\la [u(x+r)-u(x)]^4\ra$, $\la[v(x+r)-v(x)]^4\ra$, and $\la [u(x+r)-u(x)]^2[v(x+r)-v(x)]^2\ra$, where $u(x)$ is the longitudinal component of the velocity along the line joining the points ${\bf x}$ and ${\bf (x+r)}$, 
and $v(x)$ is one of the transverse components of velocity perpindicular to this line.  If these three structure functions have finite inertial ranges scaling approximately in the Kolmogorov form $ r^{4/3} $, then the integrals in Hill and Wilczak are comfortably convergent, and the pressure structure function and pressure spectrum have finite inertial ranges.  Hill and Boratav \cite{hill97} and Nelkin and Chen \cite{nel98} have recently used the Hill-Wilczak formula to calculate the pressure structure function for DNS and for experiments.  The resulting expression exhibits interesting partial cancellations, and a high degree of sensitivity to the detailed form of the fourth order velocity structure functions, but there are no singular contributions and no mathematical difficulties.

Starting from (\ref{pre}), however, great care must be taken to get a finite pressure spectrum in the limit $\nu\rightarrow 0$.  Taking the fourier transform of (\ref{pre}),  taking the absolute square, and using (\ref{diss}) and (\ref{spec}), the pressure spectrum can be written as
\begin{equation}
S_p(k) ={\frac{\la \epsilon \ra^2}{4k^5 {\nu}^2}} [C_{\epsilon} (kL)^{-\mu}+ C_D (kL)^{-\mu_{\omega}}
-F_x(kL)],
\label{cancel}
\end{equation}
where the last term on the right hand side of (\ref{cancel}) represents the cross term between enstrophy and dissipation. The most natural assumption is that the cross term scales as

\begin{equation}
F_x(kL)=C_x (kL)^{-\mu_x}
\label{cross}
\end {equation}
From a field theoretical viewpoint, it is natural to think of the enstrophy and dissipation as scaling variables with scaling dimension $\mu_{\omega}/2$ and $\mu/2$ respectively, in which case the exponent $\mu_x$ is given by

\begin{equation}
\mu_x=(\mu_{\omega}+\mu)/2
\end{equation}

 If the pressure spectrum is to be finite in the limit $\nu\rightarrow 0$, the right hand side of (\ref{cancel}), which is proportional to ${\nu}^{-2}$ must cancel identically for all values of $k$.  
If the cross term satisfies (\ref{cross}),
this is only possible if the scaling exponents in (\ref{cancel}) satisfy
\begin{equation}
{\mu}_x = {\mu}_{\omega}= \mu,
\label{exp}
\end{equation}
so that the scaling exponents for enstrophy and dissipation must be equal.

If the cross term has a more complicated structure, and does not satisfy a power law, then the argument given above no longer has a simple algebraic structure.  Taking the Fourier transform of (\ref{pre}), however, it is very hard to see how a finite value of the inertial range pressure spectrum can arise without an almost total cancellation of the two source terms on the right hand side.  Such a cancellation seems implausible unless the two source terms have the same scaling exponent.

At this stage of the argument, this cancellation of the dominant terms gives a pressure spectrum which is identically zero.  This just emphasizes that (\ref{pre}) is not a good starting point for calculating the pressure spectrum.  In fact, both (\ref{diss}) and (\ref{spec}) also have non-singular background terms which have been omitted here.  In the quasi-Gaussian approximation, these background terms will scale as ${\nu}^2 {\epsilon}^{4/3} k^{5/3}$, and will give a non-vanishing contribution to the pressure spectrum of order ${\epsilon}^{4/3} k^{-7/3}$ as expected from conventional dimensional arguments of the 1941 Kolmogorov type \cite{monin}.  In the individual spectra (\ref{diss}), and (\ref{spec}), however, these background terms are negligible in the inertial range compared to the dominant anomalous scaling terms.  It is only after the dominant singularity has been completely cancelled in the pressure that these remaining terms become important.  As discussed in \cite{hill97} and \cite{nel98} these remaining terms are sensitive to deviations from the quasi-Gaussian approximation so that the pressure spectrum and structure function are able to probe detailed properties of the fourth order velocity structure frunctions.

Finally I would add two statements about the underlying physics.  First I would emphasize that the result presented here applies only in the limit of very high Reynolds number.  As discussed in \cite{he98}, the apparent scaling can be different for finite but very large Reynolds number.  When discussing subtle questions of small differences in scaling, experiment may be far from asymptotic, even for atmospheric turbulence.  Second, I find it interesting that the basic result derived here from simple scaling arguments reminiscent of statistical field theory has been also been derived in \cite{he98} from considerations of localized vortex structures.  It is rare in turbulence theory that the same result is obtained from the statistical point of view and from the point of view of vortex structures.

The idea for this short derivation occurred at the meeting ``Turbulence: Challenges for the 21st Century," at Los Alamos, May 18-21, 1998.  This meeting was also in celebration of Robert H. Kraichnan's 70th birthday.  A key suggestion was made by Boris Shraiman, who pointed out to me at that meeting that the right hand side of (\ref{pre}) must cancel almost everywhere.  From this I have deduced,
with quite modest assumptions,  that the differences in scaling of enstrophy and dissipation observed by various authors must be a finite range scaling effect.  This is in agreement with the recent model calculation of He et al \cite{he98}, and can be thought of as a modest birthday gift from myself to Bob Kraichnan.

\end{multicols}

\end{document}